\documentclass[preprint,nofootinbib]{revtex4}%
\usepackage{amssymb}
\usepackage{amsfonts}
\usepackage{amsmath}
\usepackage{graphicx,color}
\usepackage{graphicx}%
\providecommand{\U}[1]{\protect\rule{.1in}{.1in}}
\providecommand{\U}[1]{\protect\rule{.1in}{.1in}}
\begin{document}
\title{Asymptotically warped anti-de Sitter spacetimes\\in topologically massive gravity}
\author{Marc Henneaux$^{1,2}$, Cristi\'{a}n Mart\'{\i}nez$^{1}$, Ricardo
Troncoso$^{1}$}
\email{henneaux@ulb.ac.be, martinez@cecs.cl, troncoso@cecs.cl.}
\affiliation{$^{1}$Centro de Estudios Cient\'{\i}ficos (CECs), Casilla 1469, Valdivia, }
\affiliation{$^{2}$Physique th\'{e}orique et math\'{e}matique and International Solvay
Institutes, Universit\'{e} Libre de Bruxelles, Campus Plaine C.P.231, B-1050
Bruxelles, Belgium.}

\begin{abstract}
Asymptotically warped AdS spacetimes in topologically massive gravity with
negative cosmological constant are considered in the case of spacelike
stretched warping, where black holes have been shown to exist. We provide a
set of asymptotic conditions that accommodate solutions in which the local
degree of freedom (the \textquotedblleft massive graviton\textquotedblright)
is switched on. An exact solution with this property is explicitly exhibited
and possesses a slower fall-off than the warped AdS black hole. The boundary
conditions are invariant under the semidirect product of the Virasoro algebra
with a $u(1)$ current algebra. We show that the canonical generators are
integrable and finite. When the graviton is not excited, our analysis is
compared and contrasted with earlier results obtained through the covariant
approach to conserved charges. In particular, we find agreement with the
conserved charges of the warped AdS black holes as well as with the central
charges in the algebra.


\end{abstract}
\preprint{CECS-PHY-11/05}

\maketitle

\section{Introduction}

Warped AdS spacetimes \cite{Spindel-Rooman, Warped-2, Stephane} have attracted
considerable interest recently as possible ground states of topologically
massive gravity with a negative cosmological constant
\cite{Strominger-warped-AdS}. They also become relevant in the context of
Kerr/CFT correspondence since they emerge in the near horizon geometry of
extremal Kerr black holes \cite{Kerr/CFT}.

Topologically massive gravity with a negative cosmological constant
\cite{Deser:1982vy} is described by the action \cite{Footnote}%
\begin{equation}
I[e]=2\int\ \left[  e^{a}\left(  d\omega_{a}+\frac{1}{2}\epsilon_{abc}%
\omega^{b}\omega^{c}\right)  +\frac{1}{6}\frac{1}{\ell^{2}}\epsilon_{abc}%
e^{a}e^{b}e^{c}\right]  +\frac{1}{\mu}\int\omega^{a}\left(  d\omega_{a}%
+\frac{1}{3}\epsilon_{abc}\omega^{b}\omega^{c}\right)
\end{equation}
where $\mu\neq0$ is the mass parameter, and $l$ is the AdS radius. Hereafter
we will focus on the case $|\mu l|>3$ for which black holes with spacelike
stretched warping have been shown to exist, \cite{BC,Strominger-warped-AdS}.

Asymptotically warped AdS spacetimes have already been explored in Refs.
\cite{CD1, Compere-Detournay}. The study provided there can be regarded as the
analog of the boundary analysis of \cite{Brown-Henneaux} for asymptotically
AdS spacetimes.

These analyses cover black holes. However, topologically massive gravity has
one local degree of freedom, which one can call the \textquotedblleft
graviton\textquotedblright. In order to accommodate solutions in which this
local degree of freedom is excited, it was shown in \cite{HMT-TMG-1,
HMT-TMG-2} that the boundary conditions\ of \cite{Brown-Henneaux} for
asymptotically AdS spacetimes need to be relaxed. It turns out that a similar
relaxation is also necessary in the context of asymptotically warped AdS
spacetimes. Indeed, as shown here, there exist solutions describing a
propagating massive graviton on a warped AdS black hole, and these do not
fulfill the boundary conditions of \cite{Compere-Detournay}, \cite{Relaxed
Warped}.

We provide in this paper a set of relaxed asymptotic conditions that
accommodate these solutions in which the local degree of freedom is switched
on. Just as in the absence of the graviton \cite{Compere-Detournay}, these
boundary conditions are invariant under the semidirect product of the Virasoro
algebra with a $u(1)$ current algebra. We compute the charges associated with
these symmetries and show that they are finite, integrable, and as expected
from the theorem of \cite{Brown-Henneaux-2}, they fulfill a centrally extended
version of the asymptotic symmetry algebra.

The plan of the paper is as follows. In the next Section we provide an exact
solution describing a propagating massive graviton on a warped AdS black hole.
The suitable boundary conditions that accommodate this class of solutions are
discussed in Sec. \ref{Boundary Conditions sec}. We then study the asymptotic
symmetry algebra (Section \ref{ASG sec}) and study its canonical realization
in Sec. \ref{Algebra of charges sec}. Next we study the restricted boundary
conditions obtained by switching off the graviton and we compare and contrast
them with the earlier results of \cite{Compere-Detournay} and
\cite{Blagojevic-Cvetkovic}. In particular, we find in that case agreement
with the conserved charges of the warped AdS black holes as well as with the
central charges in the algebra. The last Section is devoted to conclusions.
Some technical developments are given in the appendix.

\section{New solution: Warped AdS black hole with a bouncing graviton}

\label{Exact solution}

Our starting point are the warped AdS black holes \cite{BC}, given by%
\begin{align}
d\bar{s}^{2}  &  =dt^{2}+\frac{l^{2}}{3+\nu^{2}}\frac{dr^{2}}{(r-r_{+}%
)(r-r_{-})}-2\left(  \nu r+\frac{1}{2}\sqrt{r_{+}r_{-}(3+\nu^{2})}\right)
dtd\phi\label{Warped AdS black hole}\\
&  +\frac{r}{4}\left[  3(\nu^{2}-1)^{2}r+(3+\nu^{2})(r_{+}+r_{-})+4\nu
\sqrt{r_{+}r_{-}(3+\nu^{2})}\right]  d\phi^{2}\ ,\nonumber
\end{align}
which have been expressed in the coordinates of \cite{Strominger-warped-AdS}.
Note in particular that $\phi$ is an angle, $0\leq\phi\leq2\pi$. In
(\ref{Warped AdS black hole}), $\nu$ is equal to
\[
\nu=\frac{\mu l}{3}\ .
\]
If one sets $r_{+}=r_{-}=0$ one gets the zero mass black hole, whose metric
explicitly reads,%
\begin{equation}
ds_{0}^{2}=dt^{2}+\frac{l^{2}}{3+\nu^{2}}\frac{dr^{2}}{r^{2}}-2\nu
rdtd\phi+\frac{3}{4}(\nu^{2}-1)^{2}r^{2}d\phi^{2}\ .\label{Background}%
\end{equation}
The terms containing $r_{+}$ and $r_{-}$ are subleading so that all the
metrics of the above form share the same asymptotics.

Stretched warped AdS can be recovered from (\ref{Background}) once the
corresponding coordinate $\phi$ is unwrapped. For this reason, it is customary
to say that the solutions (\ref{Warped AdS black hole}) are asymptotic to
warped AdS although they are only so locally since $\phi$ is an angle in
(\ref{Warped AdS black hole}) while it is actually unwrapped in spacelike
stretched warped AdS spacetime. In this paper we will adopt the same abuse of language.

In order to describe a massive graviton propagating on an arbitrary warped AdS
black hole at the full non linear level, we follow the Kerr-Schild method
\cite{Debney:1969zz}. We thus look for a solution of the form%
\begin{equation}
ds^{2}=d\bar{s}^{2}+A(r,t,\phi)k_{\mu}k_{\nu}dx^{\mu}dx^{\nu}%
\ ,\label{NewMetric}%
\end{equation}
where
\[
k=k^{\mu}\partial_{\mu}\ ,
\]
is a null geodesic vector of $d\bar{s}^{2}$; i.e., $k^{\mu}k_{\mu}=0$, and
$k^{\mu}\bar{\nabla}_{\mu}k^{\nu}=0$. We choose the one-form $k=k_{\mu}%
dx^{\mu}$ to be
\begin{equation}
k=\frac{2}{3+\nu^{2}}\frac{l^{2}}{(r-r_{+})(r-r_{-})}dr+ld\phi\ .\label{Formk}%
\end{equation}
Even though $k^{\mu}$ is not a Killing vector, the field equations can be
readily integrated because they become linear in the unknown function $A$. A
particular class of solutions interesting for our purposes is given by%

\begin{equation}
\label{FunctionA}A=e^{-\omega t}\left(  \frac{\left(  r-r_{-}\right)  ^{2\nu
r_{-}+\sqrt{r_{+}r_{-}(3+\nu^{2})}}}{\left(  r-r_{+}\right)  ^{2\nu
r_{+}+\sqrt{r_{+}r_{-}(3+\nu^{2})}}}\right)  ^{\frac{\omega l}{(r_{+}%
-r_{-})(3+\nu^{2})}}F\left[  \frac{2l}{(r_{+}-r_{-})(3+\nu^{2})}\log\left(
\frac{r-r_{-}}{r-r_{+}}\right)  -\phi\right]  \ ,
\end{equation}
where $F$ is an arbitrary function, and
\begin{equation}
\omega l=\omega_{\pm}l=-3\nu\pm\sqrt{4\nu^{2}-3}\ .\label{Omega}%
\end{equation}

Note that the solution remains a solution if one simultaneously change
$t\rightarrow-t$, and $\phi\rightarrow-\phi$, a coordinate transformation that
preserves the spacetime orientation. This coordinate change amounts to making
$\omega\rightarrow-\omega$ in the solution.

A detailed physical interpretation of this solution will be given in
\cite{HMT-preprint}, where the behaviour of the metric as $t \rightarrow
\pm\infty$ and the isometries will be systematically investigated. It will
also be shown in that work that a similar solution constructed from a BTZ
black hole \cite{BTZ}\ as background \textquotedblleft seed\textquotedblright%
\ metric exists.

For the purposes of this paper, the interest of the metric (\ref{NewMetric})
with $k_{\mu}$ and $A$ given by (\ref{Formk}) and (\ref{FunctionA}),
respectively, is that it motivates the boundary conditions below. Note that
the branch $\omega=\omega_{-}$ is such that the curvature does not approach to
the one of Warped AdS at infinity. For this reason it will be discarded from
now on and we set $\omega\equiv\omega_{+}$.

\section{Boundary conditions \ }

\label{Boundary Conditions sec}

As we now show, the branch $\omega_{+}$ is asymptotically warped AdS in a
relaxed sense as compared with the results of \cite{Compere-Detournay}.

The asymptotic conditions are written with respect to the background, which is
chosen as the warped AdS black hole with $r_{+}=r_{-}=0$. Thus, the asymptotic
form of the metric is given by $ds^{2}=ds_{0}^{2}+\Delta g_{\mu\nu}dx^{\mu
}dx^{\nu}$.

We adopt as boundary conditions
\begin{equation}%
\begin{array}
[c]{lll}%
\Delta g_{rr} & = & \!\displaystyle h_{rr}^{(1)}\ r^{\alpha-4}+f_{rr}%
\ r^{-3}+Y_{rr}\ r^{2\alpha-6}+h_{rr}^{(2)}\ r^{\alpha-5}+c_{rr}\ r^{-4}%
+\cdot\cdot\cdot\\
\Delta g_{rt} & = & \!\displaystyle h_{rt}^{(1)}\ r^{\alpha-4}+f_{rt}%
\ r^{-3}+\cdot\cdot\cdot\\
\Delta g_{r\phi} & = & \!\displaystyle h_{r\phi}^{(1)}\ r^{\alpha-2}+Y_{r\phi
}\ r^{2\alpha-4}+h_{r\phi}^{(2)}\ r^{\alpha-3}+f_{r\phi}\ r^{-2}+\cdot
\cdot\cdot\\
\Delta g_{tt} & = & \!\displaystyle O(r^{-3})\\
\Delta g_{t\phi} & = & \!\displaystyle h_{t\phi}^{(1)}\ r^{\alpha-1}+f_{t\phi
}+Y_{t\phi}\ r^{2\alpha-3}+h_{t\phi}^{(2)}\ r^{\alpha-2}+c_{t\phi}%
\ r^{-1}+\cdot\cdot\cdot\\
\Delta g_{\phi\phi} & = & \!\displaystyle h_{\phi\phi}^{(1)}\ r^{\alpha
}+f_{\phi\phi}\ r+Y_{\phi\phi}\ r^{2\alpha-2}+h_{\phi\phi}^{(2)}\ r^{\alpha
-1}+c_{\phi\phi}+\cdot\cdot\cdot
\end{array}
\label{Boundary conditions-general}%
\end{equation}
where the $h^{\prime}s$, $Y^{\prime}s$, $f^{\prime}s$ and $c^{\prime}s$ are
arbitrary functions of time and the angle. $\alpha$ is an arbitrary fixed
constant which must be smaller than two, and which for simplicity we assume to
be in the range
\[
0<\alpha<\frac{3}{2}\ ,\ \alpha\neq1\ .
\]
If $\alpha$ were not in this range there would be additional terms in the
expansion. Note in particular that the cases $\alpha=0,1$ require special care
because of the potential presence of logarithmic terms.

The functions $h$'s and $Y$'s are present only when the graviton is switched
on. They are zero when it is not excited, a simplified context that will be
studied in section VI below.

The boundary conditions have been motivated by an examination of the
asymptotic behaviour of the above solution which has $\alpha=-2\nu\frac{\omega
l}{3+\nu^{2}}$, with $\omega$ given by $\omega_{+}$ in Eq. (\ref{Omega}). Note
that for the particular solution above, $\Delta g_{t\phi}$ vanishes. We have
included a nontrivial $\Delta g_{t\phi}$ in the asymptotic conditions as this
is required by the consistency of the asymptotic analysis carried out below.

The boundary conditions (\ref{Boundary conditions-general}) must be completed
by additional constraints on the coefficients appearing in the expansion in
powers of $r$ in the asymptotic form of the metric. As we will see below,
these coefficients must be subject to some conditions that ensure finiteness
of the canonical generators. These conditions are relatively intricate and for
that reason are written in the appendix. These conditions are fulfilled by the
exact solution.

We are now going to verify the consistency of the asymptotic conditions. This
involves two steps. First, the boundary conditions should be invariant under
the expected asymptotic symmetry algebra, namely the semi direct product of
the Virasoro algebra ($V$) with the $u(1)$ current algebra, i.e.,
$V\times_{\sigma}u(1)^{+}$. Second, the surface integrals appearing in the
canonical generators should be finite. We verify these crucial existence
requirements in the next two sections.

\section{Asymptotic symmetry algebra}

\label{ASG sec}

The asymptotic conditions (\ref{Boundary conditions-general}) are mapped into
themselves under the following asymptotic Killing vectors%
\begin{align}
\eta^{t}  &  =T+\frac{4\nu l^{2}}{\left(  3+\nu^{2}\right)  ^{2}}\frac{1}%
{r}\partial_{\phi}^{2}X+\cdot\cdot\cdot\nonumber\\
\eta^{\phi}  &  =X+\frac{2l^{2}}{(3+\nu^{2})^{2}}\frac{1}{r^{2}}\partial
_{\phi}^{2}X+\cdot\cdot\cdot\label{asympt-symm-warped}\\
\eta^{r}  &  =-r\ \partial_{\phi}X+\cdot\cdot\cdot\nonumber
\end{align}
where $X$ and $T$ are functions of $\phi$ only. These asymptotic
transformations not only preserve the asymptotic form of the metric, but also
the extra conditions on the coefficients that must be imposed by finiteness of
the charges, as we pointed out, and which are written in the appendix.

The asymptotic Killing vectors close in the Lie bracket according to the semi
direct product of the Virasoro algebra with the $u(1)$ current algebra,
$V\times_{\sigma}u(1)^{+}$. The Virasoro algebra is parametrized by $X(\phi)$
while the $T(\phi)$ spans the affine Kac-Moody extension of $u(1)$.
Explicitly, the commutation relations read $\left[  \eta_{1},\eta_{2}\right]
=\eta_{3}$, with%
\begin{align*}
X_{3}  &  =X_{1}\partial_{\phi}X_{2}-X_{2}\partial_{\phi}X_{1}\ ,\\
T_{3}  &  =X_{1}\partial_{\phi}T_{2}-X_{2}\partial_{\phi}T_{1}\ .
\end{align*}
As it will be seen below, the subleading terms in (\ref{asympt-symm-warped})
are crucial for getting a non vanishing central charge in the canonical
realization of the algebra.

It is worth pointing out that the subset of the asymptotic conditions
(\ref{Boundary conditions-general}) for which the coefficients $f$'s and $c$'s
are time independent is still consistent with the asymptotic symmetries. This
is because $X(\phi)$ and $T(\phi)$ depend only on $\phi$ and not on $t$, and
so do not introduce through the Lie action a time-dependence in the $f$'s and
the $c$'s (which do not mix with the $h$'s and the $Y$'s) if there is none to
begin with. Assuming the $f$'s and the $c$'s not to depend on time -- while
allowing the \textquotedblleft graviton-characterizing\textquotedblright%
\ coefficients $h$'s and $Y$'s to depend on time -- does not exclude the above
metric (\ref{NewMetric}). Therefore, this is an interesting subset of the
asymptotic conditions, which actually turns out to simplify dramatically the
formulas for the charges. These formulas are in fact explicitly given below
only in that case. This is sufficient, as our purposes are to accommodate the
graviton on an arbitrary warped AdS black hole. The extra conditions on the
metric coefficients necessary for ensuring finiteness of the charges are also
spelled out in the appendix only in that case.

\section{Canonical generators}

\label{Algebra of charges sec}

\subsection{Surface integrals}

We compute the conserved charges within the canonical formalism,
\textquotedblleft\`{a} la Regge-Teitelboim\textquotedblright%
\ \cite{Regge-Teitelboim}. As explained there, the charges that generate the
diffeomorphisms (\ref{asympt-symm-warped}) take the form
\begin{equation}
H[\eta]=\hbox{``Bulk piece"}+Q[T]+Q[X]\,,\label{generator}%
\end{equation}
where the bulk piece is a linear combination of the constraints with
coefficients involving $\eta^{t},\eta^{\phi}$, and $\eta^{r}$, and where
$Q[T]$ and $Q[X]$ are surface integrals at infinity that involve only the
asymptotic form of the vector field $\eta^{t},\eta^{\phi}$, and $\eta^{r}$. On
shell, the bulk piece vanishes and $H[\eta]$ reduces to $Q[T]+Q[X]$.

The canonical analysis of topologically massive gravity has been performed in
\cite{Canonical TMG}, and here we follow the procedure devised in
\cite{Carlip} and further developed in \cite{HMT-TMG-2}.

The general variation of the surface integrals was derived in \cite{HMT-TMG-2}%
. Applying this to the above asymptotic behaviour, one gets variations of the
surface integrals $\delta Q[T]$ and $\delta Q[X]$ which are finite thanks to
the extra conditions on the metric coefficients. These variations are
furthermore integrable without requiring further conditions.

As pointed out in the previous section, the subset of the asymptotic
conditions (\ref{Boundary conditions-general}) for which the coefficients
$f$'s and $c$'s are time independent is consistent with the asymptotic
symmetries. In this case, the corresponding charges are%
\begin{equation}
Q[T]=\frac{2}{3l\nu}\int d\phi\ T\ \left[  3(\nu^{2}-1)f_{t\phi}+2\nu
f_{\phi\phi}\right]  \ ,\label{QT-off-subset}%
\end{equation}
and%
\begin{align}
Q[X]  &  =\frac{1}{12l^{3}\nu}\frac{1}{(3+\nu^{2})^{2}}\int d\phi\ X\ \left[
-3\left(  3+\nu^{2}\right)  ^{3}\left(  \nu^{2}-1)(2\nu^{2}-3\right)
c_{rr}\right. \nonumber\\
&  +2l^{2}\left(  9\nu\left(  \nu^{2}-1\right)  \left(  3+\nu^{2}\right)
\left(  3+5\nu^{2}\right)  c_{t\phi}+2\left(  \left(  3+\nu^{2}\right)
\left(  18+3\nu^{2}+11\nu^{4}\right)  c_{\phi\phi}\right.  \right.
\label{QX-off-sub}\\
&  +3\left(  \nu^{2}-1\right)  (27-39\nu^{2}+32\nu^{4})f_{t\phi}%
^{2}\nonumber\\
&  \left.  \left.  \left.  +8\nu\left(  27-24\nu^{2}+13\nu^{4}\right)
f_{t\phi}f_{\phi\phi}+\left(  45-54\nu^{2}+25\nu^{4}\right)  f_{\phi\phi}%
^{2}\right)  \right)  \right]  \ .\nonumber
\end{align}
In these formulas we have adjusted the integration constants such that the
charges vanish for the background configuration (\ref{Background}). The
\textquotedblleft$rr$\textquotedblright\ component of the field equations
allows to express $c_{rr}$ in terms of the rest of the coefficients appearing
in $Q[X]$ (see appendix). This equation turns out to be invariant under the
asymptotic symmetries. One can verify that on-shell, $Q[X]$ simplifies to%
\begin{align}
Q[X]  &  =\frac{1}{6l\nu^{2}}\int d\phi\ X\ \left[  (5\nu^{2}+3)(3(\nu
^{2}-1)c_{t\phi}+2\nu c_{\phi\phi})\right. \nonumber\\
&  \left.  +12\nu(\nu^{2}-1)f_{t\phi}^{2}+16\nu^{2}f_{t\phi}f_{\phi\phi}+2\nu
f_{\phi\phi}^{2}\right]  \ .\label{QX-on-shelll}%
\end{align}

The same procedure applies to the general situation when the coefficients
$f$'s and $c$'s depend on time: the charges exist and are finite but their
expression is much more cumbersome.

The charges (\ref{QT-off-subset}) and (\ref{QX-off-sub}) (or
(\ref{QX-on-shelll})) possess the intriguing feature of not depending on the
coefficients $h$'s and the $Y$'s characterizing the graviton. Thus, the
graviton does not contribute directly to the asymptotic charges in
topologically massive gravity with the above boundary conditions, even though
it does enter non trivially these boundary conditions. A similar phenomenon
was encountered in the study of asymptotically AdS spacetimes in topologically
massive gravity \cite{HMT-TMG-1,HMT-TMG-2} (see also \cite{Sezgin}).

\subsection{Algebra of charges}

According to the general theorem demonstrated in \cite{Brown-Henneaux-2} the
charges close in the Poisson bracket according to the algebra of the
corresponding asymptotic symmetries, modulo possible central extensions. For
that reason, the only task for getting the algebra of the charges is to
compute the central extensions. To that end it is useful to observe that under
an asymptotic symmetry, the relevant coefficients in (\ref{QT-off-subset}) and
(\ref{QX-on-shelll}) transform as%
\begin{align*}
\delta f_{\phi\phi}  &  =f_{\phi\phi}\ \partial_{\phi}X+X\ \partial_{\phi
}f_{\phi\phi}-2\nu\ \partial_{\phi}T\ ,\\
\delta f_{t\phi}  &  =f_{t\phi}\ \partial_{\phi}X+X\ \partial_{\phi}f_{t\phi
}+\partial_{\phi}T\ ,\\
3(\nu^{2}-1)\delta c_{t\phi}+2\nu\delta c_{\phi\phi}  &  =6(\nu^{2}%
-1)c_{t\phi}\partial_{\phi}X-\frac{9}{\nu^{2}+3}X\partial_{\phi}c_{t\phi}\\
&  +\nu\left(  4c_{\phi\phi}\partial_{\phi}X+4f_{t\phi}\partial_{\phi}%
T-\frac{4}{\nu^{2}+3}l^{2}\partial_{\phi}^{3}X\right. \\
&  \left.  +X\ \left(  3\nu\frac{\nu^{2}+2}{\nu^{2}+3}\partial_{\phi}c_{t\phi
}+2\partial_{\phi}c_{\phi\phi}\right)  \right)  \,.
\end{align*}
Using these expressions we immediately find the central charge and the Poisson
bracket of the generators, which read explicitly as%
\[
\left[  Q(\eta_{1}),Q(\eta_{2})\right]  =Q(\left[  \eta_{1},\eta_{2}\right]
)+K(\eta_{1},\eta_{2})\ ,
\]
with
\[
K(\eta_{1},\eta_{2})=-\frac{2}{3l\nu}\int d\phi\ \left[  (\nu^{2}%
+3)T_{1}\partial_{\phi}T_{2}+l^{2}\frac{5\nu^{2}+3}{\nu^{2}+3}X_{1}%
\partial_{\phi}^{3}X_{2}\right]  \ .
\]

In terms of the Fourier components, the algebra reads%
\begin{align*}
i\left[  L_{m},L_{n}\right]   &  =(m-n)L_{m+n}+\frac{c_{V}}{12}\ m^{3}%
\delta_{m+n,0}\ ,\\
i\left[  L_{m},T_{n}\right]   &  =-nT_{m+n}\ ,\\
i\left[  T_{m},T_{n}\right]   &  =-\frac{c_{u(1)}}{12}\ m\delta_{m+n,0}\ ,
\end{align*}
with
\[
c_{V}=\frac{5\nu^{2}+3}{\nu(\nu^{2}+3)}\frac{l}{G}\ ;\ c_{u(1)}=\frac{\nu
^{2}+3}{\nu}\frac{l}{G}\ .
\]

\section{Switching off the graviton: restricted boundary conditions}

When the graviton is not excited, the subset of boundary conditions
(\ref{Boundary conditions-general}) further simplifies to%
\begin{equation}%
\begin{array}
[c]{lll}%
\Delta g_{rr} & = & \!\displaystyle f_{rr}\ r^{-3}+c_{rr}\ r^{-4}+\cdot
\cdot\cdot\\
\Delta g_{rt} & = & \!\displaystyle O(r^{-3})\\
\Delta g_{r\phi} & = & \!\displaystyle O(r^{-2})\\
\Delta g_{tt} & = & \!\displaystyle O(r^{-3})\\
\Delta g_{t\phi} & = & \!\displaystyle f_{t\phi}+c_{t\phi}\ r^{-1}+\cdot
\cdot\cdot\\
\Delta g_{\phi\phi} & = & \!\displaystyle f_{\phi\phi}\ r+c_{\phi\phi}%
+\cdot\cdot\cdot
\end{array}
\label{Boundary conditions-restricted}%
\end{equation}
and the condition on the coefficients that guarantees the finiteness of the
charges just reads%
\begin{equation}
\left(  3+\nu^{2}\right)  ^{2}f_{rr}-4l^{2}\left(  2\nu f_{t\phi}+f_{\phi\phi
}\right)  =0\ .\label{Condition-restricted}%
\end{equation}
It is easy to check that the warped AdS black holes
(\ref{Warped AdS black hole}) fulfill all these restricted boundary conditions
as well as the condition (\ref{Condition-restricted}) on the coefficients.

Since the charges obtained in the previous section do not depend on the $h$'s
and the $Y$'s, they take exactly the same form in this restricted setting. The
algebra of the charges, and in particular the central charges, are of course unchanged.

As stated in the introduction, this restricted setting was already
investigated in \cite{Compere-Detournay}, and \cite{Blagojevic-Cvetkovic}. As
can be checked by direct inspection, while our boundary conditions turn out to
be different, we obtain the same central charges, but our algebra only agrees
with the one in \cite{Blagojevic-Cvetkovic}. There is another important
difference between our approach and the previous treatment of
\cite{Compere-Detournay}. While this latter treatment uses the equations of
motion, and so defines the charges only for on-shell configurations, our
approach enables one to define the conserved charges even for off-shell
configurations, satisfying the asymptotic conditions
(\ref{Boundary conditions-restricted}). This appears to be important for the
path integral.

\section{Summary and final remarks}

In this paper the set of asymptotic conditions given by Eq.
(\ref{Boundary conditions-general}) has been shown to accommodate
asymptotically warped AdS solutions in the case of spacelike stretched warping
where the local degree of freedom is switched on. An explicit exact solution
of this sort, possessing a slower fall-off than the warped AdS black hole was
provided in Section \ref{Exact solution}. The relaxed set of boundary
conditions turn out to be invariant under the semidirect product of the
Virasoro algebra with a $u(1)$ current algebra. The canonical generators were
found to be integrable and finite, while the subleading terms in
(\ref{asympt-symm-warped}) were shown to play a key r\^{o}le in order to
obtain a non vanishing central charge.

Remarkably, the subset of the asymptotic conditions
(\ref{Boundary conditions-general}), for which the $f$ and the $c$
coefficients are time independent, is still consistent with the asymptotic
symmetries, which allows to simplify the form of the canonical generators as
in Eqs. (\ref{QT-off-subset}) and (\ref{QX-on-shelll}).

As it occurs for asymptotically AdS spacetimes in topologically massive
gravity \cite{HMT-TMG-1,HMT-TMG-2} the charges have the intriguing feature of
not depending on the relaxation terms switched on by the graviton
\cite{Footnote2}. Therefore, the global charges associated to the new solution
presented here, reduce to the ones of the \textquotedblleft
seed\textquotedblright\ warped AdS black hole metric
(\ref{Warped AdS black hole}), i.e.,%
\[
M=Q[\partial_{t}]=\frac{3+\nu^{2}}{24lG}\left(  r_{+}+r_{-}+\nu^{-1}%
\sqrt{r_{+}r_{-}(3+\nu^{2})}\right)  \ ,
\]
and%
\[
J=Q[\partial_{\phi}]=\nu\frac{3+\nu^{2}}{96lG}\left[  \frac{5\nu^{2}+3}%
{4\nu^{2}}(r_{+}-r_{-})^{2}-\left(  r_{+}+r_{-}+\nu^{-1}\sqrt{r_{+}r_{-}%
(3+\nu^{2})}\right)  ^{2}\right]  \ .
\]
This agrees with the results of \cite{Compere-Detournay} and
\cite{Blagojevic-Cvetkovic} for the mass and the angular momentum of the
warped AdS black hole in the absence of the graviton \cite{Footnote 3}.

A similar construction can be performed for asymptotically AdS spacetimes in
topologically massive gravity, where an analog solution constructed out from a
BTZ black hole as background metric can be shown to exist. Further details
about this solution, and about the similar effect of non-appearance of the
graviton-characterizing asymptotic coefficients in the charges, will be
discussed in \cite{HMT-preprint}.

\bigskip

\textit{Acknowledgments.} We thank G. Comp\`{e}re, S. Detournay and D. Tempo,
for useful discussions. C. M. and R. T. wish to thank the kind hospitality at
the Physique th\'{e}orique et math\'{e}matique at the Universit\'{e} Libre de
Bruxelles and the International Solvay Institutes. M. H. gratefully
acknowledges support from the Alexander von Humboldt Foundation through a
Humboldt Research Award and support from the ERC through the \textquotedblleft
SyDuGraM\textquotedblright\ Advanced Grant. This research is partially funded
by FONDECYT grants N$%
{{}^\circ}%
$ 1085322, 1095098, 1100755, and by the Conicyt grant ACT-91:
\textquotedblleft Southern Theoretical Physics Laboratory\textquotedblright%
\ (STPLab). The work of MH is also partially supported by IISN - Belgium
(conventions 4.4511.06 and 4.4514.08) and by the Belgian Federal Science
Policy Office through the Interuniversity Attraction Pole P6/11. The Centro de
Estudios Cient\'{\i}ficos (CECs) is funded by the Chilean Government through
the Centers of Excellence Base Financing Program of Conicyt.

\appendix

\section{Conditions for the finiteness of the charges}

In this appendix, we consider only the subset of the asymptotic conditions
(\ref{Boundary conditions-general}), for which the $f^{\prime}s$ and
$c^{\prime}s$ coefficients are time independent. This is done for the sole
purpose of simplicity, in order to get formulas which, even though formidable,
remain tractable. The general case can be treated along similar lines but is
even more formidable.

The coefficients appearing in the expansion in powers of $r$ in the asymptotic
form of the metric (\ref{Boundary conditions-general}) are subject to certain
conditions that ensure the finiteness of the canonical generators. This can be
seen as follows. For asymptotic conditions of the form
(\ref{Boundary conditions-general}) the corresponding charges spanned by the
asymptotic symmetries (\ref{asympt-symm-warped}) contain a finite piece, given
by $Q[T]$ and $Q[X]$ in Eqs. (\ref{QT-off-subset}) and (\ref{QX-off-sub}),
respectively, as well as some terms that diverge as $r\rightarrow\infty$
according to
\begin{align}
Q[T,X]  &  =Q[T]+Q[X] +\frac{1}{8l^{3}\nu\left(  3+\nu^{2}\right)  }\int
d\phi\ \left[  X \left(  \frac{r^{\alpha}}{3}\ \mathcal{X}_{(\alpha)}
-(\nu^{2}-1)(5\nu^{2}-3)\ r\mathcal{X}_{(1)} \right.  \right. \nonumber\\
&  \left.  \left.  +\frac{ r^{2\alpha-2}\mathcal{X}_{(2\alpha-2)}+
r^{\alpha-1} \mathcal{X}_{(\alpha-1)} }{3l^{2}\left(  3+\nu^{2}\right)  }
\right)  +\frac{4 T\ r^{\alpha-1}}{3 }\mathcal{T}_{(\alpha-1)}\right]
.\label{Q+Divs}%
\end{align}
Therefore, the canonical generators become finite for an arbitrary asymptotic
symmetry provided $\mathcal{X}_{(i)}$ and $\mathcal{T}_{(\alpha-1)}$ vanish.

The simplest condition is independent of $\alpha$ and corresponds to the
vanishing of the divergence that is linear in $r$, i.e., $\mathcal{X}_{(1)}=0
$ which only involves the $f^{\prime}s$ and it reduces to Eq.
(\ref{Condition-restricted}). The leading divergence, $r^{a}$, gives a linear
condition that depends on the $h^{(1)\prime}s$ and their derivatives, which
reads%
\begin{align}
&  \mathcal{X}_{(\alpha)}=-3\left(  \nu^{2}-1\right)  \left(  3+\nu
^{2}\right)  ^{2}\left(  3(-2+\alpha)+(4+\alpha)\nu^{2}\right)  h_{rr}%
^{(1)}\nonumber\\
&  +l^{2}\left(  12\nu\left(  \nu^{2}-1\right)  \left(  9+15\nu^{2}%
+\alpha(-7+2\alpha)\left(  3+\nu^{2}\right)  \right)  h_{t\phi}^{(1)}\right.
\\
&  +4\left(  9(-2+\alpha)^{2}+6(1+\alpha(-8+3\alpha))\nu^{2}+(22+\alpha
(-12+5\alpha))\nu^{4}\right)  h_{\phi\phi}^{(1)}\nonumber\\
&  \left.  \left.  +3\left(  \nu^{2}-1\right)  (-4(-3+2\alpha)\nu\left(
3+\nu^{2}\right)  \partial_{t}h_{r\phi}^{(1)}+3(\nu^{2}-1)(\nu^{2}%
+3)\partial_{t}^{2}h_{rr}^{(1)}-4l^{2}\partial_{t}^{2}h_{\phi\phi}%
^{(1)}\right)  \right)  \ .\nonumber
\end{align}
There are two conditions coming from the divergence that go like $r^{\alpha
-1}$, given by%
\begin{align}
\mathcal{T}_{(\alpha-1)}  &  =6\nu^{3}\left(  3+\nu^{2}\right)  ^{2}%
h_{rr}^{(1)}-6l^{2}(2\nu^{2}(3+5\nu^{2})+\alpha(3+\nu^{2})(3-5\nu^{2}%
+\alpha(\nu^{2}-1)))h_{t\phi}^{(1)}\nonumber\\
&  -4l^{2}\nu\left(  6\nu^{2}+\alpha(3+\nu^{2}\right)  (\alpha-3))h_{\phi\phi
}^{(1)}+6l^{2}\left(  3+\nu^{2}\right)  \left(  \alpha(\nu^{2}-1)-2\nu
^{2}\right)  \partial_{t}h_{r\phi}^{(1)}\\
&  -3l^{2}\nu(\nu^{2}-1)(\nu^{2}+3)\partial_{t}^{2}h_{rr}^{(1)}+4l^{4}%
\nu\partial_{t}^{2}h_{\phi\phi}^{(1)}\ ,\nonumber
\end{align}
involving only and the $h^{(1)\prime}s$ and their derivatives, as well as a
nonlinear one that reads
\begin{align}
\mathcal{X}_{(\alpha-1)}  &  =\left(  3+\nu^{2}\right)  ^{2}f_{rr}\left(
3\left(  -1+\nu^{2}\right)  \left(  3+\nu^{2}\right)  ^{2}\left(
6(-3+\alpha)+(9+2\alpha)\nu^{2}\right)  h_{rr}^{(1)}\right. \nonumber\\
&  -l^{2}\left(  6\nu\left(  -1+\nu^{2}\right)  \left(  3+37\nu^{2}%
+\alpha(-13+4\alpha)\left(  3+\nu^{2}\right)  \right)  h_{t\phi}^{(1)}\right.
\nonumber\\
&  +2\left(  9(8+\alpha(-9+2\alpha))+6(-3+\alpha(-13+6\alpha))\nu
^{2}+(58+\alpha(-17+10\alpha))\nu^{4}\right)  h_{\phi\phi}^{(1)}\nonumber\\
&  \left.  \left.  -6(-5+4\alpha)\nu\left(  -3+2\nu^{2}+\nu^{4}\right)
\partial_{t}h_{r\phi}^{(1)}+9\left(  -1+\nu^{2}\right)  ^{2}\left(  3+\nu
^{2}\right)  \partial_{t}^{2}h_{rr}^{(1)}\right)  \right) \nonumber\\
&  +l^{2}\left(  2f_{\phi\phi}\left(  3\left(  -1+\nu^{2}\right)  \left(
3+\nu^{2}\right)  ^{2}\left(  3\alpha+(-18+\alpha)\nu^{2}\right)  h_{rr}%
^{(1)}\right.  \right. \nonumber\\
&  -2l^{2}\left(  2\nu\left(  9(-23+5\alpha)+6(5+(35-8\alpha)\alpha)\nu
^{2}+(-207+(65-16\alpha)\alpha)\nu^{4}\right)  h_{t\phi}^{(1)}\right.
\nonumber\\
&  -4\left(  -18(-2+\alpha)+3(-9+2\alpha(-5+2\alpha))\nu^{2}+(47+4(-2+\alpha
)\alpha)\nu^{4}\right)  h_{\phi\phi}^{(1)}\nonumber\\
&  \left.  \left.  +2\nu\left(  3+\nu^{2}\right)  \left(  -27+(-21+16\alpha
)\nu^{2}\right)  \partial_{t}h_{r\phi}^{(1)}+3\left(  9+9\nu^{2}-13\nu
^{4}-5\nu^{6}\right)  \partial_{t}^{2}h_{rr}^{(1)}+16l^{2}\nu^{2}\partial
_{t}^{2}h_{\phi\phi}^{(1)}\right)  \right) \nonumber\\
&  +4f_{t\phi}\left(  3\nu\left(  -1+\nu^{2}\right)  \left(  3+\nu^{2}\right)
^{2}\left(  3(3+\alpha)+(-15+\alpha)\nu^{2}\right)  h_{rr}^{(1)}\right. \\
&  +2l^{2}\left(  3\left(  -1+\nu^{2}\right)  \left(  9\left(  6+\alpha
-\alpha^{2}\right)  +6(-2+\alpha(-14+3\alpha))\nu^{2}+(110+\alpha
(-29+7\alpha))\nu^{4}\right)  h_{t\phi}^{(1)}\right. \nonumber\\
&  +2\nu\left(  99-18\alpha^{2}+12(-6+(-3+\alpha)\alpha)\nu^{2}%
+(85+6(-2+\alpha)\alpha)\nu^{4}\right)  h_{\phi\phi}^{(1)}\nonumber\\
&  -3\left(  -3(2+\alpha)+(-8+7\alpha)\nu^{2}\right)  \left(  -1+\nu
^{2}\right)  \left(  3+\nu^{2}\right)  \partial_{t}h_{r\phi}^{(1)}\nonumber\\
&  \left.  \left.  +9\nu\left(  -1+\nu^{2}\right)  ^{2}\left(  3+\nu
^{2}\right)  \partial_{t}^{2}h_{rr}^{(1)}-12l^{2}\nu\left(  -1+\nu^{2}\right)
\partial_{t}^{2}h_{\phi\phi}^{(1)}\right)  \right) \nonumber\\
&  -3\left(  -1+\nu^{2}\right)  \left(  3+\nu^{2}\right)  ^{3}\left(
3(-3\ +\alpha)+(3+\alpha)\nu^{2}\right)  h_{rr}^{(2)}\nonumber\\
&  +12l^{2}\nu\left(  -1+\nu^{2}\right)  \left(  3+\nu^{2}\right)  \left(
36+24\nu^{2}+\alpha(-11+2\alpha)\left(  3+\nu^{2}\right)  \right)  h_{t\phi
}^{(2)}\nonumber\\
&  +4l^{2}\left(  3+\nu^{2}\right)  \left(  9(-3+\alpha)^{2}+6(12+\alpha
(-14+3\alpha))\nu^{2}+(39+\alpha(-22+5\alpha))\nu^{4}\right)  h_{\phi\phi
}^{(2)}\nonumber\\
&  -8l^{2}(-3+\alpha)\left(  3+\nu^{2}\right)  ^{2}\left(  3+5\nu^{2}\right)
\partial_{\phi}h_{r\phi}^{(1)}-12l^{2}(-5+2\alpha)\nu\left(  -1+\nu
^{2}\right)  \left(  3+\nu^{2}\right)  ^{2}\partial_{t}h_{r\phi}%
^{(2)}\nonumber\\
&  -18l^{2}(-4+\alpha)\left(  \left(  -1+\nu^{2}\right)  \left(  3+\nu
^{2}\right)  \right)  ^{2}\partial_{t}h_{tr}^{(1)}+12l^{2}\nu\left(
-1+\nu^{2}\right)  \left(  3+\nu^{2}\right)  ^{2}\partial_{t}\partial_{\phi
}h_{rr}^{(1)}\nonumber\\
&  +24l^{4}\left(  -1+\nu^{2}\right)  \left(  3(-2+\alpha)+(2+\alpha)\nu
^{2}\right)  \partial_{t}\partial_{\phi}h_{t\phi}^{(1)}+16l^{4}\nu\left(
2\alpha\left(  3+\nu^{2}\right)  -3\left(  7+\nu^{2}\right)  \right)
\partial_{t}\partial_{\phi}h_{\phi\phi}^{(1)}\nonumber\\
&  \left.  +9l^{2}\left(  9-12\nu^{2}-2\nu^{4}+4\nu^{6}\right)  \partial
_{t}^{2}h_{rr}^{(2)}-12l^{4}\left(  -1+\nu^{2}\right)  \left(  3+\nu
^{2}\right)  \left(  \partial_{t}^{2}h_{\phi\phi}^{(2)}+2\partial_{t}%
^{2}\partial_{\phi}h_{r\phi}^{(1)}\right)  \right) .\nonumber
\end{align}

The remaining condition, coming from the divergence of order $O(r^{2\alpha
-2})$, is a combination of linear terms in the $Y^{\prime}s$ and quadratic
involving the $h^{(1)\prime}s$ and it reads%
\begin{align}
&  \mathcal{X}_{(2\alpha-2)}=\frac{3}{2}\left(  \nu^{2}-1\right)  \left(
3+\nu^{2}\right)  ^{4}\left(  -24+7\nu^{2}+4\alpha\left(  3+\nu^{2}\right)
\right)  (h_{rr}^{(1)})^{2}\nonumber\\
&  -24l^{2}\left(  \nu^{2}-1\right)  \left(  3+\nu^{2}\right)  ^{2}\left(
-6+\nu^{2}+\alpha\left(  3+\nu^{2}\right)  \right)  (h_{r\phi}^{(1)}%
)^{2}\nonumber\\
&  +12l^{4}\left(  \nu^{2}-1\right)  \left(  63+84\nu^{2}+141\nu^{4}%
-76\alpha\nu^{2}\left(  3+\nu^{2}\right)  +\alpha^{2}\left(  -9+66\nu
^{2}+23\nu^{4}\right)  \right)  (h_{t\phi}^{(1)})^{2}\nonumber\\
&  +8l^{4}\left(  72+15\nu^{2}+57\nu^{4}+2\alpha^{2}\left(  9+24\nu^{2}%
+7\nu^{4}\right)  -4\alpha\left(  18+27\nu^{2}+7\nu^{4}\right)  \right)
(h_{\phi\phi}^{(1)})^{2}\nonumber\\
&  -6l^{2}\left(  \nu^{2}-1\right)  \left(  3+\nu^{2}\right)  ^{3}\left(
-6+\nu^{2}+\alpha\left(  3+\nu^{2}\right)  \right)  Y_{rr}\nonumber\\
&  +12l^{4}\nu\left(  -3+2\nu^{2}+\nu^{4}\right)  \left(  75+37\nu
^{2}-30\alpha\left(  3+\nu^{2}\right)  +8\alpha^{2}\left(  3+\nu^{2}\right)
\right)  Y_{t\phi}\nonumber\\
&  +8l^{4}\left(  3+\nu^{2}\right)  \left(  72+87\nu^{2}+33\nu^{4}%
-8\alpha\left(  9+15\nu^{2}+4\nu^{4}\right)  +2\alpha^{2}\left(  9+18\nu
^{2}+5\nu^{4}\right)  \right)  Y_{\phi\phi}\nonumber\\
&  -12l^{2}\nu\left(  \nu^{2}-1\right)  \left(  3+\nu^{2}\right)  ^{3}%
h_{r\phi}^{(1)}\partial_{t}h_{rr}^{(1)}-\frac{9}{2}l^{2}\left(  \nu
^{2}-1\right)  ^{2}\left(  3+\nu^{2}\right)  ^{3}(\partial_{t}h_{rr}%
^{(1)})^{2}\nonumber\\
&  +6l^{2}(-7+6\alpha)\nu\left(  \nu^{2}-1\right)  \left(  3+\nu^{2}\right)
^{3}h_{rr}^{(1)}\partial_{t}h_{r\phi}^{(1)}+12l^{4}\left(  \nu^{2}-1\right)
\left(  3+\nu^{2}\right)  ^{2}(\partial_{t}h_{r\phi}^{(1)})^{2}\nonumber\\
&  +12l^{4}\left(  -3+2\nu^{2}+\nu^{4}\right)  \left(  -2\left(  15-7\nu
^{2}+2\alpha\left(  -3+\nu^{2}\right)  \right)  h_{r\phi}^{(1)}+3\nu\left(
\nu^{2}-1\right)  \partial_{t}h_{rr}^{(1)}\right)  \partial_{t}h_{t\phi}%
^{(1)}\nonumber\\
&  -2l^{4}\left(  -8\nu\left(  3+\nu^{2}\right)  \left(  -39-\nu^{2}%
+3\alpha\left(  7+\nu^{2}\right)  \right)  h_{r\phi}^{(1)}\right. \\
&  -\left.  3\left(  \nu^{2}-1\right)  \left(  3\left(  -3+2\nu^{2}+\nu
^{4}\right)  \partial_{t}h_{rr}^{(1)}-8l^{2}\nu\partial_{t}h_{t\phi}%
^{(1)}\right)  \right)  \partial_{t}h_{\phi\phi}^{(1)}\nonumber\\
&  -16l^{6}\left(  -3+\nu^{2}\right)  (\partial_{t}h_{\phi\phi}^{(1)}%
)^{2}-9L^{2}\left(  \nu^{2}-1\right)  ^{2}\left(  3+\nu^{2}\right)  ^{3}%
h_{rr}^{(1)}\partial_{t}^{2}h_{rr}^{(1)}\nonumber\\
&  -6l^{2}\left(  -3+2\nu^{2}+\nu^{4}\right)  h_{t\phi}^{(1)}\left(
\nu\left(  3+\nu^{2}\right)  \left(  15+41\nu^{2}-19\alpha\left(  3+\nu
^{2}\right)  +6\alpha^{2}\left(  3+\nu^{2}\right)  \right)  h_{rr}%
^{(1)}\right. \nonumber\\
&  +\left.  4l^{2}\left(  \left(  -2\left(  6+5\nu^{2}\right)  +\alpha\left(
3+9\nu^{2}\right)  \right)  \partial_{t}h_{r\phi}^{(1)}-3\nu\left(  \nu
^{2}-1\right)  \partial_{t}^{2}h_{rr}^{(1)}\right)  \right) \nonumber\\
&  +2l^{2}h_{\phi\phi}^{(1)}\left(  -3\left(  3+\nu^{2}\right)  ^{2}\left(
-10\alpha\left(  3+4\nu^{2}+\nu^{4}\right)  +\alpha^{2}\left(  9+18\nu
^{2}+5\nu^{4}\right)  +2\left(  12+\nu^{2}+11\nu^{4}\right)  \right)
h_{rr}^{(1)}\right. \nonumber\\
&  +2l^{2}\left(  2\nu\left(  2\alpha^{2}\left(  -9+66\nu^{2}+23\nu
^{4}\right)  +3\left(  87+26\nu^{2}+79\nu^{4}\right)  -\alpha\left(
81+402\nu^{2}+125\nu^{4}\right)  \right)  h_{t\phi}^{(1)}\right. \nonumber\\
&  +\left.  \left.  \left(  3+\nu^{2}\right)  \left(  \nu\left(
90-36\alpha+(54-44\alpha)\nu^{2}\right)  \partial_{t}h_{r\phi}^{(1)}+3\left(
-3-2\nu^{2}+5\nu^{4}\right)  \partial_{t}^{2}h_{rr}^{(1)}\right)  \right)
\right) \nonumber\\
&  -24l^{4}\left(  \nu^{2}-1\right)  \left(  3+\nu^{2}\right)  ^{2}h_{r\phi
}^{(1)}\partial_{t}^{2}h_{r\phi}^{(1)}-32l^{6}\nu\left(  3\left(  \nu
^{2}-1\right)  h_{t\phi}^{(1)}+2\nu h_{\phi\phi}^{(1)}\right)  \partial
_{t}^{2}h_{\phi\phi}^{(1)}\nonumber\\
&  +9l^{4}\left(  -3+2\nu^{2}+\nu^{4}\right)  ^{2}\partial_{t}^{2}%
Y_{rr}-12l^{6}\left(  -3+2\nu^{2}+\nu^{4}\right)  \partial_{t}^{2}Y_{\phi\phi
}\nonumber\\
&  -12 l^{4}\nu\left(  3+\nu^{2}\right)  ^{2} \left(  -1+\nu^{2}\right)
\left(  4\alpha-7 \right)  \partial_{t} Y_{r \phi}.\nonumber
\end{align}

\bigskip

Remarkably, the asymptotic symmetries (\ref{asympt-symm-warped}) not only
preserve the asymptotic form of the metric, but also these additional
conditions on the coefficients in the expansion of the metric that are imposed
by finiteness of the charges. Indeed, all but one of the conditions written
above automatically fulfill%
\[
\mathcal{L}_{\eta}\mathcal{X}_{(i)}=0\ ;\ \mathcal{L}_{\eta}\mathcal{T}%
_{(\alpha-1)}=0\ .
\]
A nontrivial consistency condition only comes from $\mathcal{L}_{\eta
}\mathcal{X}_{(\alpha-1)}$, that reads%
\begin{align}
0  &  =3\left(  -1+\nu^{2}\right)  \left(  3+\nu^{2}\right)  ^{2}\left(
3+5\nu^{2}-3\alpha\left(  3+\nu^{2}\right)  +\alpha^{2}\left(  3+\nu
^{2}\right)  \right)  h_{rr}^{(1)}\nonumber\\
&  +l^{2}\left(  4(2-\alpha)\left(  3+\nu^{2}\right)  \left(  3+5\nu
^{2}-3\alpha\left(  3+\nu^{2}\right)  +\alpha^{2}\left(  3+\nu^{2}\right)
\right)  h_{\phi\phi}^{(1)}\right. \nonumber\\
&  \left.  -3\left(  -1+\nu^{2}\right)  \left(  3\left(  -3+2\nu^{2}+\nu
^{4}\right)  \partial_{t}^{2}h_{rr}^{(1)}-4l^{2}(-2+\alpha)\partial_{t}%
^{2}h_{\phi\phi}^{(1)}\right)  \right)  ,
\end{align}
which nevertheless may be easily verified to be also invariant under the
asymptotic symmetry group.

Besides, the \textquotedblleft$rr$\textquotedblright\ component of the field
equations that allows to simplify $Q[X]$ as in Eq. (\ref{QX-on-shelll}) is
also invariant under the asymptotic symmetries, and reads%

\begin{align}
&  c_{rr}=\frac{2 l^{2} \left(  3+5 \nu^{2}\right)  c_{t\phi}}{\nu\left(
3+\nu^{2}\right)  ^{2}}+\frac{1}{4 l^{2} \left(  -3+\nu^{2}\right)  \left(
3+\nu^{2}\right)  ^{3}}\left(  16 l^{4} \left(  -9+\nu^{4}\right)  c_{\phi
\phi}\right. \nonumber\\
&  + \left.  3 \left(  -1+\nu^{2}\right)  \left(  3+\nu^{2}\right)  ^{4}
f_{rr}^{2}-24 l^{2} \left(  -1+\nu^{2}\right)  \left(  3+\nu^{2}\right)  ^{2}
f_{rr} (2 \nu f_{t\phi}+f_{\phi\phi})\right. \\
&  \left.  -16 l^{4} \left(  -3 \left(  3-20 \nu^{2}+9 \nu^{4}\right)
f_{t\phi}^{2}+\left(  7 \nu^{2}-15\right)  f_{\phi\phi}(4 \nu f_{t\phi} +
f_{\phi\phi}) \right)  \right) .\nonumber
\end{align}

It is straightforward to verify that all of the conditions spelled above are
fulfilled by the new exact solution in Section \ref{Exact solution}.

\bigskip

As stated above, we have derived in this appendix the extra constraints on the
coefficients in the metric expansion only for the subset of the asymptotic
conditions (\ref{Boundary conditions-general}) where the $f^{\prime}s$ and
$c^{\prime}s$ coefficients are time independent. A tedious, but
straightforward computation shows that when the $f^{\prime}s$ and $c^{\prime
}s$ depend on time, the boundary conditions (\ref{Boundary conditions-general}%
) are still consistent with the asymptotic symmetries in the sense explained
here, but with even more intricate finiteness and consistency conditions.


\begin{thebibliography}{99}
\bibitem {Footnote}Here we are using $16\pi G=1$, a metric of signature
$(-++)$, and a cosmological constant $\Lambda=-1/l^{2}$, and set
$\epsilon_{012}=1$ (so $\epsilon^{012}=-1$).
\bibitem {Relaxed Warped}Asymptotically warped AdS solutions possesing a
relaxed behavior as compared with the one in \cite{Compere-Detournay} can also
be found in \cite{Grumiller-warped, Chow-Pope-Sezgin}, as well as in
\cite{Dio-Monica-Mboyo} at the linearized level.
\end{thebibliography}
\end{document}